\documentclass{ws-ijmpe}
\begin{document}
\markboth{M. G\'o\'zd\'z and W. A. Kami\'nski}{Localization on Fat
Branes as the Source of Neutrino Mixing}
\title{LOCALIZATION ON FAT BRANES \\ AS THE SOURCE OF NEUTRINO MIXING}
\author{MAREK G\'O\'ZD\'Z and WIES{\L}AW A. KAMI\'NSKI}
\address{
Department of Theoretical Physics, Maria Curie-Sk{\l}odowska University, \\ 
Radziszewskigo 10, PL-20-031 Lublin, Poland \\ 
mgozdz@kft.umcs.lublin.pl \\
kaminski@neuron.umcs.lublin.pl}
\maketitle
\begin{abstract}
  The localization of fermions in extra dimensions, proposed by
  Arkani-Hamed and Schmaltz, is discussed as the source of the
  phenomenon of particle mixing. We work out the example of neutrinos in
  detail.
\end{abstract}
\bigskip

The possibility of existence of additional dimensions is an old idea and
has been extensively studied in the literature (see
e.g.\cite{mgozdz:extraD}). Only recently this topic received a serious
back up from the string theory, which requires at least six new spatial
dimensions. The usual way is to assume that the additional dimensions
are very small, therefore till now undetectable. During last few years
Arkani-Hamed, Dimopoulos and Dvali (ADD) suggested that this does not
need to be true.\cite{mgozdz:add} The so-called ``large extra
dimensions'' have been shown to explain the gauge hierarchy problem and
the mass hierarchies between families to some extent. In these models we
assume that our world is confined to a 3-brane, which means that all
standard model (SM) particles are constrained to propagate within a
hypersurface, embedded in higher-dimensional space (the bulk). Only
gravity is permitted to occupy the bulk as well, which resolves the
gauge hierarchy problem.

One of the possible extensions of ADD model has been proposed in
Ref.\cite{mgozdz:as}, where the brane has a non-zero width in the extra
dimension. This allows to localize the SM particles at different points
in the extra dimension, therefore separating them spatially. The main
motivation for such a setup is that by careful adjusting the distances
between the particles one obtains suppressions in interactions and may
solve the mass hierarchy problem. The next step is to accommodate the
mixing between particles (like quarks and neutrinos) in this
picture. This problem has been partially addressed in,
e.g.,\cite{mgozdz:local} for the case of linear extra dimension.

In the present paper we show, using the newest neutrino data, that the
neutrino mixing may be explained by the ADD model with fat brane in the
shape of a hyper-tube. We leave the issue of the mass hierarchy as well
as more detailed discussion of the shape of extra dimension to a
forthcoming paper.


Let us denote the coordinates by $\{x^\mu,y\}$, $\mu=1\dots4$, where $y$
corresponds to the 5th spatial dimension. We assume the shape of the
fermion wave function to be
\begin{equation}
  \Psi(x^\mu,y) = \psi(x^\mu) \chi(y),
\end{equation}
with $\psi$ being the usual Dirac spinor. For the extra-dimension
function $\chi$ we postulate a Gaussian shape:
\begin{equation}
  \chi(y) = \frac{1}{\sigma\sqrt{2\pi}} 
  \exp\left(-\frac{(y-y_0)^2}{2\sigma^2}\right),
\end{equation}
localized around the point $y_0$ in the extra dimension; $\sigma$
corresponds to the width of the function. The Gaussian seems to be the
most natural candidate for $\chi$ but, in principle, other functions are
possible.

It is a common practice to compactify the extra dimensions, although
there were many attempts to discuss particle masses and mixings assuming
the extra dimension to be an interval.\cite{mgozdz:local} In such a
case, however, there may be additional effects occuring at the ends of
the interval, like for example tunneling or leaking of the fields from
the brane to the bulk. To avoid such situations we assume that the extra
dimension forms a circle, i.e. we identify $y+L \equiv y+2\pi R \sim y$.

The mixing between two particles comes in our model from non-zero
overlaps of Gaussians in the extra dimension. The probability of
transition from a particle $a$ to particle $b$ (or from $b$ to $a$), is
proportional to the overlap squared and is given by
\begin{equation}
  Prob(a \leftrightarrow b) = \frac{1}{2\pi(\sigma_a^2 + \sigma_b^2)} 
  \exp\left(-\frac{(y_a - y_b)^2}{\sigma_a^2 + \sigma_b^2}\right).
\label{mgozdz:prob}
\end{equation}

In the case of neutrinos, their mixing is usually described by
$ |\nu_i\rangle = U_{ij} |\nu_j\rangle,$ where $i=e,\mu,\tau$ labels
the flavor states and $j=1,2,3$ labels the mass eigenstates. The
standard parametrization of the unitary matrix $U$ in terms of three
mixing angles is, neglecting possible phases,
\begin{eqnarray}
  \left (
  \begin{array}{ccc}
    c_{12} c_{13} & s_{12} c_{13} & s_{13} \\
    -s_{12} c_{23} - c_{12} s_{23} s_{13} & c_{12} c_{23} - s_{12}
    s_{23} s_{13} & s_{23} c_{13} \\
    s_{12} s_{23} - c_{12} c_{23} s_{13} & -c_{12} s_{23} - s_{12}
    c_{23} s_{13} & c_{23} c_{13}
  \end{array}
  \right )
\end{eqnarray}
where $s_{ij} = \sin\theta_{ij}$, $c_{ij} = \cos\theta_{ij}$, and
$\theta_{ij}$ are the mixing angles between the eigenstates labeled by
indices $i$ and $j$. The recent global analysis of neutrino
oscillations\cite{mgozdz:maltoni} yields the best fit values: $0.23 <
{\sin}^2\theta_{12} < 0.39$, $0.31 < {\sin}^2\theta_{23} < 0.72$,
${\sin}^2\theta_{13} < 0.054$, which correspond to the following mixing
matrix squared
\begin{equation}
  |U|^2 =  
  \left (
  \begin{array}{ccc}
    (0.58-0.77) & (0.23-0.36) & (0.00-0.05) \\
    (0.16-0.23) & (0.08-0.53) & (0.31-0.67) \\
    (0.07-0.18) & (0.24-0.55) & (0.26-0.69)
  \end{array}
  \right ).
\label{mgozdz:uk}
\end{equation}
Each entry of $|U|^2$ represents the probability of finding neutrino of
one state in another state, thus we have
\begin{equation}
  (|U|^2)_{ij} = Prob(\nu_i \leftrightarrow \nu_j),
\end{equation}
which should be understood as a set of nine equations, with the LHS
taken from Eq. (\ref{mgozdz:uk}) and the RHS having the form of
Eq. (\ref{mgozdz:prob}). We have analyzed numerically this set of
equations, obtaining values of possible $\sigma$'s and $y$'s.

%
\begin{figure*}
  \includegraphics[width=1.0\textwidth]{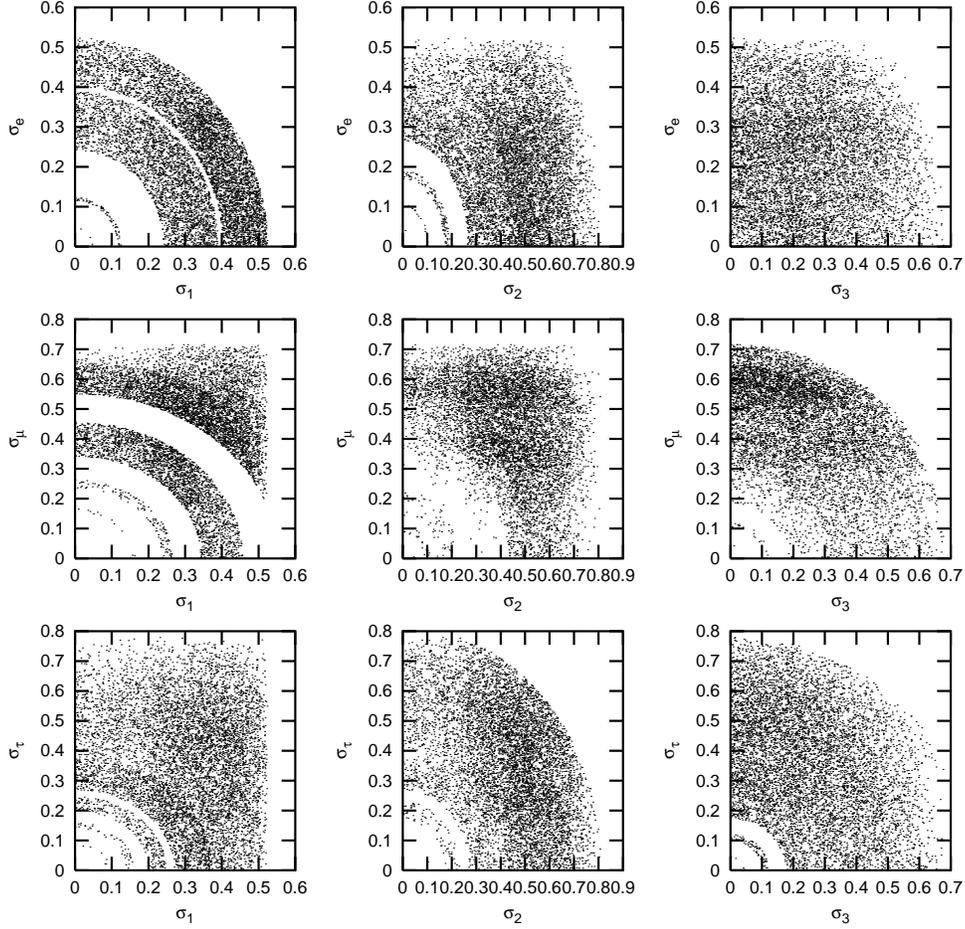}
  \caption{\label{mgozdz:sigma} Allowed values of the $\sigma$
  parameters for $L=2$.}
\end{figure*}
\begin{figure}[!h]
  \includegraphics[width=0.9\textwidth]{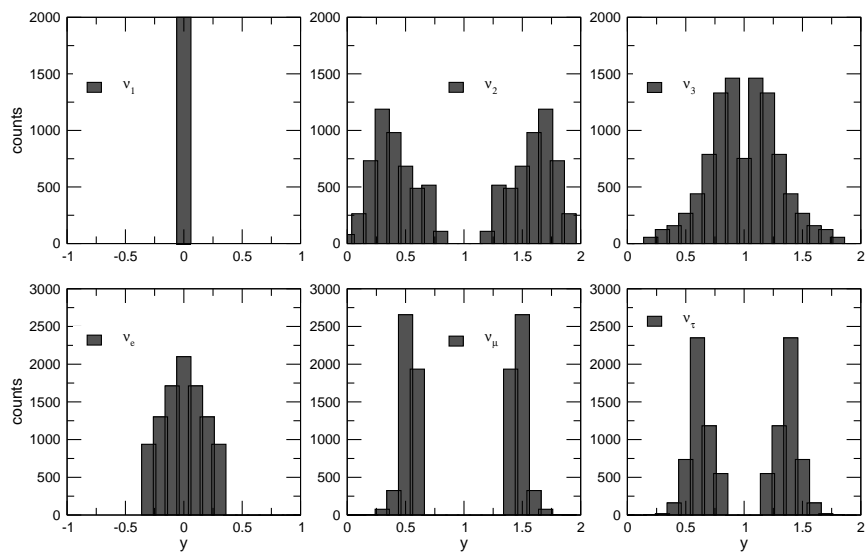}
  \caption{\label{mgozdz:y} Possible localizations of neutrinos for
  $L=2$.}
\end{figure}

%
The results for the widths are presented on Fig.~\ref{mgozdz:sigma}. The
disallowed regions come from the form of Eq.~(\ref{mgozdz:prob}), for
which a logarithm will appear in the solutions. The widths of the
allowed regions depend on the uncertainty in entries of $|U|^2$. Despite
being in some cases difficult to recognize, the same pattern emerges on
all nine panels of Fig.~\ref{mgozdz:sigma}. The corresponding
localizations of the Gaussians are shown on Fig.~\ref{mgozdz:y}. Here,
we have fixed the localization of $\nu_1$ at $y=0$ on the circle. The
length of the compactified extra dimension was set to 2 in some
arbitrary units of length. The symmetric pattern is present because of
the mirror symmetry in localizations, which will show up in the cases of
all periodic shapes (it means simply that we can count the distance
going clockwise or anticlockwise along the circle). This figure does not
show the detailed alignment of the fields. In order to give a full
presentation one should find all the patterns which show up in the
results and classify them. This is, however, beyond the scope of the
present paper. One general feature is immediately visible, namely that
there are preferred locations for each neutrino type. We have checked,
that it is independent of the length of the circle.

In summary, we have shown that the observed neutrino mixing pattern may
have its source in the geometry of our world. The not mentioned here
mass hierarchy among neutrinos finds also an explanation in our picture.

This work is supported by the Polish State Committee for Scientific
Research under grants no.~2P03B~071~25 and 1P03B~098~27. The authors
would like to thank Prof. A. Faessler for his warm hospitality in
T\"ubingen during the Summer 2004.

\pagebreak

\end{document}